\newcommand{\cal}{\mathcal}
\newcommand{\ud}{\mathrm{d}}
\newcommand{\dm}{\displaystyle}
\newcommand{\uN}{{\mathrm{N}}}
\newcommand{\uZ}{{\mathrm{Z}}}
\newcommand{\sss}{\scriptscriptstyle}
\newcommand{\sus}{\scriptstyle}
\newcommand{\m}{{\sus-}}

\documentclass[epsfig,amsmath,amssymb,showpacs]{elsart}
\usepackage{amsmath}
\usepackage{amssymb}
\usepackage{epsfig}
\usepackage{graphicx}
\usepackage{epic}
\usepackage{eepic}
\usepackage{dcolumn}
\usepackage{bm}
\usepackage{rotating}

\begin{document}
\begin{frontmatter}
\title{A model for consecutive spallation and fragmentation reactions in inverse kinematics
at relativistic energies}

\author[GSI,IPN]{P. Napolitani},
\author[IPN]{L. Tassan-Got},
\author[GSI]{P. Armbruster},
\author[IPN]{M. Bernas}		

\address[GSI]{GSI, Planckstr. 1, 64291 Darmstadt, Germany}
\address[IPN]{IPN Orsay, IN2P3, 91406 Orsay, France}

\begin{abstract}
Secondary reactions induced by relativistic beams 
in inverse kinematics in a thick target 
are relevant in several fields of experimental 
physics and technology, like secondary radioactive beams, 
production of exotic nuclei close to the proton drip line, 
and cross-section measurements for applications of spallation
reactions for energy production and incineration of nuclear wastes.
A general mathematical formulation is presented and successively
applied as a tool to disentangle the primary reaction yields from
the secondary production in the measurement of fission of a
$^{238}$U projectile impinging on a proton target at the 
energy of 1 A GeV.
\newline
{\it keywords:} 
SECONDARY AND MULTIPLE REACTIONS IN INVERSE KINEMATICS; 
In-flight identification in Z and A by magnetic spectrometer, 
Measured primary spallation and fragmentation cross sections;
Fission and evaporation residues cross sections;
Spallation reaction, p($^{238}$U,X), E=1 AGeV; 
Relevance for the production of radioactive beams.			      
\end{abstract}
\end{frontmatter}
\thispagestyle{plain}

\section{Introduction}

The reaction residues of projectile spallation and fragmentation, produced with 
heavy-ion beams can undergo consecutive nuclear reactions in the target.
The secondary reaction production, especially by the use of a thick
target, can extend towards exotic nuclei, which would not
be generated in the primary reaction.
The induction of multiple reactions in thick targets is also a
technique to obtain exotic beams.
On the other hand, in the measurements dedicated to the extraction of
formation cross sections, the secondary reactions enter in
competition with the primary production and are a disturbing
process.
In this work we will refer to a recent measurement of the
isotopic production cross section in the reaction 
$^{238}$U+p at 1 A GeV \cite{Bernas02,Taieb98,Ricciardi02}.
The experiment was performed in inverse kinematics with the
high-resolution magnetic spectrometer FRS \cite{FRS} (GSI, Darmstadt) directing
the uranium beam on a hydrogen target.
The target thickness was optimised in order to maximize the primary
production rate in respect to parasitic reactions 
and to limit the average share of secondary reactions to values of about 
$5\%
$.
This share and its uncertainty enters in the total experimental
uncertainty;
though, some of the measured isotopes, suffer from a much larger
competition of the secondary reactions.

When the primary isotopic production extends over a wide 
neutron-number range, as it will be pointed out,
we can not extract reliable cross sections in the neutron-deficient
side.
Especially for the fission residues, the approaching towards the
residue corridor defines a limit for the measurement technique: in this case the
neutron-deficient isotopes will reveal to be mostly or entirely
produced by secondary spallation of fission residues.

The beginning of the present work is dedicated to the derivation of
the ma\-the\-ma\-ti\-cal formalism to treat consecutive reactions in a
general form, independently of any application.
Furthermore, we will focus on the cross-section measurement, and an
analytical recipe to disentangle the secondary contribution from the
primary production is presented.
The method has been successfully applied in the data analysis of the
reaction $^{238}$U+p at 1 A GeV.

\section{Derivation of the exact formulation 
for multiple reactions}
%
\subsection{Attenuation of the beam}
 
When a beam of ions characterised by atomic number $\uZ_0$ and neutron
number $\uN_0$ interacts with a layer of matter, the intensity 
${\cal I}_0$ evolves accordingly to the total reaction cross-section
of the projectile $\sigma_0 = \sigma(\uN_0\uZ_0)$ and the properties of
the target, like the density $\rho\,[g\,cm^{-3}]$ and the mass of the
target nuclei $A_{\mathrm{tar}}\,[g]$.
The loss $-\mathrm{d}{\cal I}_0(\xi)$ of the beam intensity is
proportional to the path $\mathrm{d}\xi$ traversed inside of the target
medium and to the intensity ${\cal I}_0(\xi)$ reached at the position $\xi$
in the beam direction:
\[
	-\mathrm{d}{\cal I}_0(\xi)\,=\,
	a_{\sus \mathrm{tar}}\sigma_0
	\cdot
	{\cal I}_0(\xi)
	\mathrm{d}\xi
	.
\]
The target properties are represented by 
$a_{\mathrm{tar}}\,[cm^{-1}mb^{-1}]$ $=$ 
$10^{-27} \cdot {\cal N_{\sss A}}\cdot  \rho / A_{\mathrm{tar}}$,
with ${\cal N_{\sss A}}$ indicating the Avogadro's number.
Integrating over the whole path $x$ we obtain the probability for the
projectile $(\uN_0\uZ_0)$ to survive for a path $x$ in the layer of
matter as
\begin{equation}
	{\cal P}_0(\uN_0\uZ_0,\chi)
	\,=\,
	{\cal I}_0(x)/{\cal I}_0(0)
	\,=\, 
	e^{\dm -\sigma_0 a_{\sus \mathrm{tar}} x}
	\,=\, 
	e^{\dm -\sigma_0\chi}
	,
\label{eq:0order}
\end{equation}
where $\chi \,[mb^{-1}]=a_{\sus \mathrm{tar}} x$ 
indicates the number of target atoms 
per $10^{-27} cm^2 $.

\subsection{Probability of primary reaction}

The probability for the projectile $(\uN_0\uZ_0)$ to interact with 
the target in a path length $\mathrm{d}\xi$ with a total reaction 
cross-section $\sigma_0$ and produce a residue $(\uN_1\uZ_1)$ 
with a probability $p_{(0\to 1)}$ is defined as
\[
	\mathrm{d}{P}(\uN_0\uZ_0,\uN_1\uZ_1,\mathrm{d}\xi)
	\,=\,
	a_{\sus \mathrm{tar}}\mathrm{d}\xi
	\cdot
	\sigma_0
	\cdot 
	p_{(0\to 1)}
	\,=\,
	\mathrm{d}\zeta
	\cdot
	\sigma_{(0\to 1)}	
	,
\]
where $\sigma_{(0\to 1)} = \sigma_0\cdot p_{(0\to 1)}$ results to be the 
production cross
section for the reaction $(\uN_0\uZ_0 \to \uN_1\uZ_1)$,
and $\mathrm{d}\zeta = a_{\sus \mathrm{tar}}\mathrm{d}\xi$
is the number of atoms  in a layer of matter defined by the path 
$\mathrm{d}\xi$ per $10^{-27} cm^2 $.
The total probability for the projectile $(\uN_0\uZ_0)$ to react 
only once in the target in any position $\xi$ and produce the observed 
fragment $(\uN_1\uZ_1)$ is expressed by the probability
for the projectile to survive for a path $\xi$, react in $\mathrm{d}\xi$
and produce a residue that will traverse the remaining length of the
target without any further reaction; as pictured in the drowing below, 
this expression should be integrated over
any path $\xi$ in the form :
\begin{equation}\begin{split}
	&
	\int_{0}^{x}
	\left[
	{\cal P}_0(\uN_0\uZ_0,\xi)
	\cdot
	\mathrm{d}{P}(\uN_0\uZ_0,\uN_1\uZ_1,d\xi)
	\cdot
	{\cal P}_0(\uN_1\uZ_1,x\m \xi)
	\right]
	\,=
	\\&\qquad=
	\sigma_{(0\to1)}e^{\dm -\sigma_1\chi}
	\int_{0}^{\chi}
	{\cal P}_0(\uN_0\uZ_0,\zeta)
	\,
	e^{\dm \sigma_1\zeta} 
	\,
	\ud \zeta
	\notag
	.
\end{split}\end{equation}
\setlength{\unitlength}{1mm}
\begin{center}\begin{picture}(80,18)(0,15)
	\linethickness{2pt}
	\thicklines
	\put(10,26){\makebox(0,0)[b]{$\uN_0\uZ_0$}}
	\put(70,26){\makebox(0,0)[b]{$\uN_1\uZ_1$}}
	\put(20,16){\makebox(0,0)[b]{$0$}}
	\put(41,16){\makebox(0,0)[b]{$\xi$}}
	\put(60,16){\makebox(0,0)[b]{$x$}}
	\put(41,31){\makebox(0,0)[b]{$\leftrightarrow$}}
	\put(46,31.5){\makebox(0,0)[b]{$\ud \xi$}}
	\linethickness{1pt}
	\thicklines
	\dashline[+80]{2}(10,25)(38,25)
	                    \put(38,25){\line(-2,1) {1}}
	                    \put(38,25){\line(-2,-1){1}}
	\dashline[+40]{2}(44,25)(70,25)
	                    \put(70,25){\line(-2,1) {1}}
	                    \put(70,25){\line(-2,-1){1}}
	\linethickness{1pt}
	\thicklines
	\put(20,20){\line(0,1){10}}
	\put(60,20){\line(0,1){10}}
	\put(20,20){\line(1,0){40}}
	\put(20,30){\line(1,0){40}}
	\put(41,25){\circle*{0.8}}
	\put(41,25){\circle{2}}
	\thinlines\put(41,25){\circle{3}}
	\thinlines\dashline[+30]{1}(42,20)(42,30)
	\thinlines\dashline[+30]{1}(40,20)(40,30)
\end{picture}\end{center}
Introducing Eq.~(\ref{eq:0order}) in the integral,
the solution expressed in terms of target thickness gives:
\begin{equation}
	{\cal P}_1(\uN_0\uZ_0,\uN_1\uZ_1,\chi) 
	\, =\,
	-\sigma_{(0\to1)}
	\left(\frac{e^{\dm -\sigma_0\chi}}{\sigma_0-\sigma_1}+
	\frac{e^{\dm -\sigma_1\chi}}{\sigma_1-\sigma_0}\right)
	.
\label{eq:1order}
\end{equation}

\subsection{Secondary reactions}

After the primary reaction of the beam occurred, we should
consider the possibility for a further interaction of the
residues with the target. 
In this case, in order to obtain the probability of observing a
secondary fragment $\uN_2\uZ_2$, the quantity to integrate over 
any path through the target in beam direction is the product of 
three terms: the probability for the beam to undergo a primary
reaction during the path $\xi$ and produce the intermediate
fragment $\uN_1\uZ_1$, the probability for a further
reaction $(\uN_1\uZ_1 \to \uN_2\uZ_2)$ in the interval 
$\mathrm{d}\xi$, and the probability that the final residue
survives for the remaining path $x-\xi$. This description is
represented below in the drawing and it results into the following
equation:
\begin{equation}\begin{split}
	&
	\int_{0}^{x}
	\bigl[
	{\cal P}_1(\uN_0\uZ_0,\uN_1\uZ_1,\xi)
	\cdot 
	\mathrm{d}{P}(\uN_1\uZ_1,\uN_2\uZ_2,d\xi)
	\cdot
	{\cal P}_0(\uN_2\uZ_2,x\m \xi)
	\bigr]
	\,=
	\\&\qquad=
	\dm \sigma_{(1\to2)}
	\,
	e^{\dm -\sigma_2\chi}
	\int_{0}^{\chi}
	{\cal P}_1(\uN_0\uZ_0,\uN_1\uZ_1,\zeta)
	\,
	e^{\dm \sigma_2\zeta} 
	\,
	\ud \zeta
	\notag
	.
\end{split}\end{equation}
\setlength{\unitlength}{1mm}
\begin{center}\begin{picture}(80,18)(0,15)
	\linethickness{2pt}
	\thicklines
	\put(10,26){\makebox(0,0)[b]{$\uN_0\uZ_0$}}
	\put(40,26){\makebox(0,0)[b]{$\uN_1\uZ_1$}}
	\put(70,26){\makebox(0,0)[b]{$\uN_2\uZ_2$}}
	\put(20,16){\makebox(0,0)[b]{$0$}}
	\put(51,16){\makebox(0,0)[b]{$\xi$}}
	\put(60,16){\makebox(0,0)[b]{$x$}}
	\put(51,31){\makebox(0,0)[b]{$\leftrightarrow$}}
	\put(56,31.5){\makebox(0,0)[b]{$\ud \xi$}}
	\linethickness{1pt}
	\thicklines
	\dashline[+80]{2}(10,25)(28,25)
	                    \put(28,25){\line(-2,1) {1}}
	                    \put(28,25){\line(-2,-1){1}}
	\dashline[+60]{2}(34,25)(48,25)
	                    \put(48,25){\line(-2,1) {1}}
	                    \put(48,25){\line(-2,-1){1}}
	\dashline[+40]{2}(54,25)(70,25)
	                    \put(70,25){\line(-2,1) {1}}
	                    \put(70,25){\line(-2,-1){1}}
	\linethickness{1pt}
	\thicklines
	\put(20,20){\line(0,1){10}}
	\put(60,20){\line(0,1){10}}
	\put(20,20){\line(1,0){40}}
	\put(20,30){\line(1,0){40}}
	\put(51,25){\circle*{0.8}}
	\put(51,25){\circle{2}}
	\put(31,25){\circle*{0.8}}
	\put(31,25){\circle{2}}
	\thinlines\put(51,25){\circle{3}}
	\thinlines\put(31,25){\circle{3}}
	\thinlines\dashline[+30]{1}(52,20)(52,30)
	\thinlines\dashline[+30]{1}(50,20)(50,30)
\end{picture}\end{center}
Introducing Eq.~(\ref{eq:1order}) in the integral 
we obtain for the secondary reactions:
\begin{equation}\begin{split}
	&
	{\cal P}_2(\uN_0\uZ_0,\uN_1\uZ_1,\uN_2\uZ_2,\chi) 
	=\\&\quad= 
	\dm \sigma_{(0\to1)}\sigma_{(1\to2)}
	\biggl[\dm
	\frac{e^{\dm -\sigma_0\chi}}{(\sigma_0\m\sigma_1)(\sigma_0\m\sigma_2)}
	+
	\frac{e^{\dm -\sigma_1\chi}}{(\sigma_1\m\sigma_0)(\sigma_1\m\sigma_2)}
	+
	\frac{e^{\dm -\sigma_2\chi}}{(\sigma_2\m\sigma_0)(\sigma_2\m\sigma_1)}
	\biggr]
	.
\label{eq:2order}
\end{split}\end{equation}

As it will be demonstrated in the section \ref{app:norder}, 
the procedure followed so far to obtain the probability 
for secondary reactions can be extended to higher orders 
according to the recursive relation:
\begin{equation}\begin{split}
	&{\cal P}_n(\uN_0\uZ_0,\uN_1\uZ_1,\cdots,\uN_n\uZ_n,\chi) 
	\,=\\ &\quad=\, 
	\dm \sigma_{(n-1\to n)}
	e^{\dm -\sigma_n\chi}
	\int_{0}^{\chi}
	{\cal P}_{n-1}(\uN_0\uZ_0,\uN_1\uZ_1,\cdots,\uN_{n-1}\uZ_{n-1},\zeta)
	\, 
	e^{\dm \sigma_n\zeta} 
	\, 
	\ud \zeta
	.
\label{eq:recursive}
\end{split}\end{equation}
Repeating the iterative integral for higher and higher orders
we obtain the solution for the $n^{\mathrm{th}}$order:
\begin{equation}
	{\cal P}_n(\uN_0\uZ_0,\uN_1\uZ_1,\cdots,\uN_n\uZ_n,\chi) 
	\, =\, 
	(-1)^n
	\,
	\underset{\sss i=1}{\overset{\sss n}{\textstyle \prod}}
	\sigma_{(i-1\to i)}
	\,
	\sum_{i=0}^{n}
	\frac
	{e^{\dm -\sigma_i\chi}}{
	\underset{\sss j=0\atop j\ne i}{\overset{\sss n}{\prod}}
	(\sigma_i-\sigma_j)}
	.
\label{eq:norder}
\end{equation}

\section{A more stable approximated formulation}
%
If we neglect the technical and chemical properties of the
target, i.e. we consider the target homogeneous and a full
acceptance of the spectrometer, 
the relation~(\ref{eq:norder}) is formally rigorous 
and exact. 
Nevertheless, it should be observed that the term 
$(\sigma_i-\sigma_j)$ could generate instability in numerical calculations:
when very few mass is removed in one step of the chain of consecutive reactions,
the total cross-sections $\sigma_i$ and $\sigma_j$ become almost identical,
and the denominator $(\sigma_i-\sigma_j)$ becomes very small.
In this case, the sum in Eq.~(\ref{eq:norder}) remains finite but its numerical computation 
can be problematic in this form.

A possibility to remove 
this inconvenience is to perform a
series of consecutive $1^{\mathrm{st}}$-order approximations in
iterating the integration~(\ref{eq:recursive}).
We start applying the approximation to the
Eq.~(\ref{eq:1order}), which describes the probability of
primary reaction:
\begin{equation}
	{\cal P}_1(\uN_0\uZ_0,\uN_1\uZ_1,\chi) 
	\, = \,
	-\sigma_{(0\to 1)}
	\frac
	{e^{\dm -\sigma_0\chi} - e^{\dm -\sigma_1\chi}}
	{\sigma_0 - \sigma_1}
	\notag
	.
\end{equation}
Expanding the exponential to the $2^{\mathrm{nd}}$ order
in $(\sigma_0-\sigma_i)\chi$, we obtain
the simpler relation:
 \begin{equation}
	{\cal P}_1(\uN_0\uZ_0,\uN_1\uZ_1,\chi) 
	\, \approx \,
	\sigma_{(0\to1)}
	\,
	\chi
	\,
	e^{\dm -\frac{\sigma_0+\sigma_1}{2} \chi}
	.
\label{eq:1orderapprox}
\end{equation}
Integrating Eq.~(\ref{eq:1orderapprox}) by the 
relation~(\ref{eq:recursive}) we obtain the probability of
observing a secondary product generated by a given
intermediate fragment in the following form:
\begin{equation}
	\sigma_{(0\to 1)}
	\sigma_{(1\to 2)}
	\,
	e^{\dm -\sigma_2 \chi}
	\int_{0}^{\chi}
	\zeta
	e^{\dm -\left( \frac{\sigma_0+\sigma_1}{2} - \sigma_2 \right) \zeta}
	\ud \zeta
	\notag
	.
\end{equation}
Expanding to the $1^{\mathrm{st}}$ order in 
$\frac{1}{2}\left(\sigma_0+\sigma_1\right)\chi - \sigma_2\chi$,
integrating, and isolating $\chi^2 / 2$, we obtain the approximated 
form of the Eq.~(\ref{eq:2order}) in the $1^{\mathrm{st}}$ order :
\begin{equation}\begin{split}
	&
	{\cal P}_2(\uN_0\uZ_0,\uN_1\uZ_1,\uN_2\uZ_2,\chi) 
	\, \approx \,
	\sigma_{(0\to 1)}
	\sigma_{(1\to 2)}
	\,
	\frac{\chi^2}{2}
	\,
	e^{\dm -\frac{\sigma_0+\sigma_1+\sigma_2 }{3}\,\chi}
	.
\label{eq:2orderapprox}
\end{split}\end{equation}
Physically, this relation can be intuitively pictured as
``one-third-target approximation'', as the argument of the exponential
represents the attenuation of the beam, the primary and the
secondary residue, respectively, when they cross a third of the
target.

Iterating the integration~(\ref{eq:recursive}) followed 
by the first order expansion in 
respect to the argument of the exponential,
we obtain the
approximation for $n$ consecutive reactions, equivalent to
dividing the target in $n$ portions of equal thickness, each
one originating a successive reaction with a probability 
$\sigma_{(i-1\to i)}\chi/i$:
\begin{eqnarray}
	{\cal P}_n(\uN_0\uZ_0,\uN_1\uZ_1,\cdots,\uN_n\uZ_n,\chi) 
	\,&=&\,
	\underset{\sss i=1}{\overset{\sss n}{\textstyle \prod}}
	\left(
	\frac{\sigma_{(i-1\to i)}\;\chi}{i}
	\right)
	\,
	e^{\dm -
	\frac{
	{\textstyle \sum_{j=0}^{n}}
	\, \sigma_j 
	}
	{n+1}\,\chi
	}
	\notag\\
	&=&\,
	\underset{\sss i=1}{\overset{\sss n}{\textstyle \prod}}
	\sigma_{(i-1\to i)}
	\,
	\frac{\chi^n}{n!}
	\,
	e^{\dm -
	\frac{
	{\textstyle \sum_{j=0}^{n}}
	\, \sigma_j 
	}
	{n+1}\,\chi
	}
	.
\label{eq:norderapprox}
\end{eqnarray}
To obtain the Eq.~(\ref{eq:norderapprox}) we can follow the 
prescription presented in the section \ref{app:approx}.

\section{Measured reaction probability}
Besides the production rate in the target expressed in 
Eq.~(\ref{eq:norder}) , the measured production rate of an isotope NZ 
should also take into account the probability that the residue NZ
is observed outside of the target, and is detected in the spectrometer. 
The measured probability to produce NZ can be written for a given 
chain of intermediate products as
\begin{equation}
	{\cal M}_n(\uN_0\uZ_0,\cdots,\uN_n\uZ_n,\chi) 
	\,=\,
	{\cal P}_n(\uN_0\uZ_0,\cdots,\uN_n\uZ_n,\chi) 
	\cdot t_n(\uN_1\uZ_1,\cdots,\uN_n\uZ_n) 	
	\notag
	,
\end{equation}
where $t_n$ is the transmission coefficient depending on the
kinematics of all the intermediate isotopes: it depends on the instrumental
device and it represents the probability that a fragment 
with a given velocity is transmitted through the spectrometer.
The transmission is a key parameter in the measurement of fission products 
which are only partially accepted, as some of the residues are emitted
with large angle and hit the pipe of the spectrometer.
The measured probability to obtain NZ by multiple reactions 
accounts for the sum of the orders of the reactions,
that is the superposition of the contributions 
of primary, secondary and multiple reactions evaluated for any
possible choice of intermediate fragments
\begin{equation}
	{\cal M}(\uN_0\uZ_0,\uN\uZ,\chi) 
	\,=\,
	\sum_{\mathrm{order}\,n}
	\,\sum_{n-1\,\mathrm{chain}}
	{\cal M}_n(\uN_0\uZ_0,\cdots,\uN_n\uZ_n,\chi) 
	\notag
	,
\end{equation}
where the first sum accounts for the order of reactions and the 
second sum accounts for an order given 
for all the possible chains of successive $n-1$ intermediate
fragments leading from the initial projectile $\uN_0\uZ_0$ to the observed 
residue NZ.
We can express the whole relation in terms of cross-sections if we introduce 
the apparent cross section
\begin{equation}
	\tilde{\sigma}_{(\uN_0\uZ_0 \to \uN\uZ)} (\chi) 
	=
	\frac{1}{\chi}
	\frac{{\cal M}(\uN_0\uZ_0,\uN\uZ,\chi)}{t(\uN\uZ)}
	\notag
	,
\end{equation}
where $t(\uN\uZ)$ is the transmission factor of the spectrometer.
It is derived from the measured spectrum of longitudinal velocities,
by assuming the emission isotropic in respect to the centre of mass;
this assumption allows to evaluate the angular distribution of
the fragments and the fraction which is 
selected by the angular acceptance.
Due to the difference in the distributions of the emission 
velocities related to fission fragments 
or evaporation residues, respectively , the coefficient $t(\uN\uZ)$
depends strongly on the dominant reaction process
responsible for the production of a given isotope NZ.
In order to have a more complete description of the 
reaction process, we can disentangle fission 
residues from evaporation residues. 
If we assume that fission could occur only once
in a chain of successive reactions we obtain the general description:
\begin{align}
	\tilde{\sigma}_{(\uN_0\uZ_0 \to \uN\uZ)} (\chi) 
	\,\,=\,\,
	\frac{1}{\chi}
	\,
	\sum_{\mathrm{order}\,n}
	\,\,
	\sum_{n-1\,\mathrm{chain}}
	\,
	\left\{
	\left[
	{\cal R}_n^{\sus \mathrm{evr}}{\cal T}_n^{\sus \mathrm{evr}}+ 
	\sum_{j=N_1Z_1}^{j=NZ}
	\left(
	{\cal R}_n^{\sus \mathrm{evr},\atop\sus \mathrm{fis}j}
	{\cal T}_n^{\sus \mathrm{evr},\atop\sus \mathrm{fis}j} 
	\right)	
	\right]
	\,
	{\cal A}_n
	\right\}
	,
\label{eq:apparent}
\end{align}
with:
\begin{equation}
	{\cal R}_n^{\sus \mathrm{evr}} 
	\,=\,
	\frac{1}{n!}\,
	\prod_{i=\uN_1\uZ_1}^{i=\uN\uZ}
	(\sigma_{i-1\to i}^{\sus \mathrm{evr}}\chi)
	\qquad , \qquad
	{\cal R}_n^{\sus \mathrm{evr},\atop\sus \mathrm{fis}j} 
	\,=\,
	\frac{1}{n!}\,
	\sigma_{j-1\to j}^{\sus \mathrm{fis}}\chi
	\prod_{\sus i=\uN_1\uZ_1\atop \sus i\ne j}^{i=\uN\uZ}
	(\sigma_{i-1\to i}^{\sus \mathrm{evr}}\chi)
	\notag
	,
\end{equation}
\begin{equation}\begin{split}
	{\cal T}_n^{\sus \mathrm{evr}} 
	&\,=\,
	\frac{t_n^{\sus \mathrm{evr}}
	(\uN_1\uZ_1,\uN_2\uZ_2,\cdots,\uN_{n-1}\uZ_{n-1},\uN\uZ)}
	{t(\uN\uZ)}
	,
	\\
	{\cal T}_n^{\sus \mathrm{evr},\atop\sus \mathrm{fis}j} 
	&\,=\,
	\frac{t_n^{\sus \mathrm{evr},\atop\sus \mathrm{fis}j}
	(\uN_1\uZ_1,\uN_2\uZ_2,\cdots,\uN_{n-1}\uZ_{n-1},\uN\uZ)}
	{t(\uN\uZ)}
	\notag
	,
\end{split}\end{equation}
\begin{equation}\begin{split}
	{\cal A}_n(\uN_0\uZ_0,\uN_1\uZ_1,\cdots,\uN_n\uZ_n,\chi) 
	\,&=\, (-1)^n 
	\,n!\,
	\sum_{i=0}^{n}
	\frac{e^{\dm -\sigma_i\chi}}
	{
	\underset{\sss j=0\atop j\ne i}{\overset{\sss n}{\prod}}
	(\sigma_i\chi-\sigma_j\chi)}
	\,
	\approx
	\,
	e^{\dm -
	\frac{
	{\textstyle \sum_{j=0}^{n}}
	\, \sigma_j\chi 
	}
	{n+1}
	}
	\notag
	.
\end{split}\end{equation}
The first sum describes the apparent cross-section as a composition of
consecutive orders of multiple reactions.
The second sum over all the possible intermediate fragments can be
represented by an expansion in $(n-1)$ chains, the $i^{\mathrm{th}}$ one
satisfying the condition 
$(\uN_{i-1}+\uZ_{i-1})\ge (\uN_{i}+\uZ_{i})\ge (\uN_{i+1}+\uZ_{i+1})$:
each successive intermediate fragment should in fact have lower mass but
not necessarily lower charge due to the charge exchange. Nevertheless,
since this process has very low cross section, the condition can be safely
decomposed into $\uN_{i-1}\ge \uN_{i}\ge \uN_{i+1}$ and 
$\uZ_{i-1}\ge \uZ_{i}\ge \uZ_{i+1}$ for each sum $i$.
${\cal R}_n^{\sus \mathrm{evr}}$ and
${\cal R}_n^{\sus \mathrm{evr},\,\sus \mathrm{fis}j}$
are the terms defining the reaction chain:
${\cal R}_n^{\sus \mathrm{evr}}$ 
is the term representing the series of
reactions producing NZ
as a chain of consecutive evaporation residues;
${\cal R}_n^{\sus \mathrm{evr},\,\sus \mathrm{fis}j}$
differs from the former one because at the step $j$ a fission reaction
occours. 
Due to the very low fissility of the fission residues, 
the probability to undergo two consecutive fission reactions is negligible:
therefore, we imposed only one possible fission event in the chain.
The terms
${\cal T}_n^{\mathrm{evr}}$ and
${\cal T}_n^{\sus \mathrm{evr},\,\sus\mathrm{fis}j}$
are transmission coefficients ratios and depend on the optics of the
instrumental device; a more detailed description of these coefficients
will follow in the next paragraph.
${\cal A}_n$ is the term describing the attenuation of each nucleus
traversing the target, either the beam-like projectile and the consecutive
residues. It should be observed that this terms takes into consideration
also the attenuation of the final residue, due to the probability to
undergo a reaction of order $n+1$; nevertheless no reaction of order
$n+1$ appears in the term describing the reaction chain.
As discussed in the previous paragraph, according to the possibility to
have an unstable solution, it is important to decide in which condition the 
exact form containing the term $\sigma_i - \sigma_j$ and referred to in 
Eq.~(\ref{eq:norder}) can be applied;
it could be safer to use the approximated form represented by the 
Eq.~(\ref{eq:norderapprox}) 
which remains valid as long as the products $(\sigma_i-\sigma_j)\chi$ 
are small compared to unity.
\begin{figure*}[b]
\includegraphics[angle=-90,width=1\textwidth]{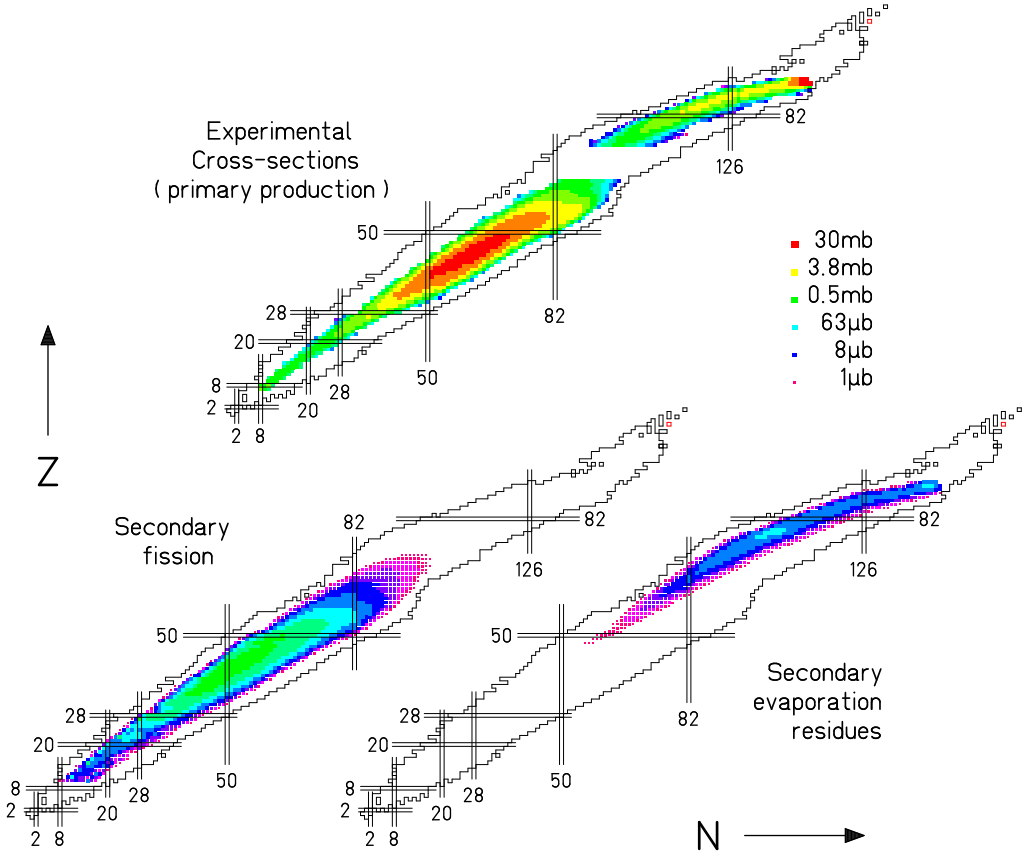}
\caption{\label{fig:secondary}
	Top.
Experimental cross-sections for the reaction
$^{238}$U+p at 1 A GeV \cite{Bernas02,Taieb98,Ricciardi02} measured
in inverse kinematics and corrected for the secondary contribution. 
A hydrogen target of thickness $87.3 \, mg/cm^2$
was used. 
	Bottom.
Contribution of the secondary reactions to the appearent cross section
as calculated with Eq.~(\ref{eq:apparent}). 
The secondary production is
more than one order of magnitude lower than the primary production.
The secondary fission residues, shown on the left, 
account for primary evaporation residues
followed by secondary fission residues
or primary fission residues followed by secondary
evaporation residues.
The primary evaporation residues followed by secondary evaporation residues 
are shown on the right.
}
\end{figure*}

We simulated the secondary and ternary production (expressed as apparent
cross-sections) in the reaction of $^{238}$U at $1$A GeV 
impinging on a target of hydrogen of 
$87.3\,10^{-3} [\mathrm{g}/\mathrm{cm}^2]$
by solving the system~(\ref{eq:apparent}) with the choice of the approximated 
form~(\ref{eq:norderapprox}). 
The production cross-sections 
$\sigma_{\uN_0\uZ_0\to \uN_i\uZ_i}^{\sus \mathrm{evr}}$ and
$\sigma_{\uN_0\uZ_0\to \uN_i\uZ_i}^{\sus \mathrm{fis}}$
were measured experimentally in inverse kinematics at the FRagment Separator 
in Darmstadt \cite{Bernas02,Taieb98,Ricciardi02}. The total cross-sections $\sigma_i$ have been 
calculated numerically according to the model of Karol-Brohm \cite{Karol75,Brohm94}, very
well suited for proton reactions. Another choice would be the
formula of Benesh et al. \cite{Benesh89}, where the total cross-section is calculated neglecting
the energy of the projectile and the nuclear properties of the target: as a
consequence of these approximations the description of ion-proton reactions
gives an almost-systematic over-prediction of around 20\% in the total cross
section; nevertheless, in the case of ion-ion reactions, the formula of 
Benesh et al. gives satisfactory results and is then preferable due to the lower 
computing time.
\begin{figure}[b]
\begin{center}
\mbox{\epsfig{file=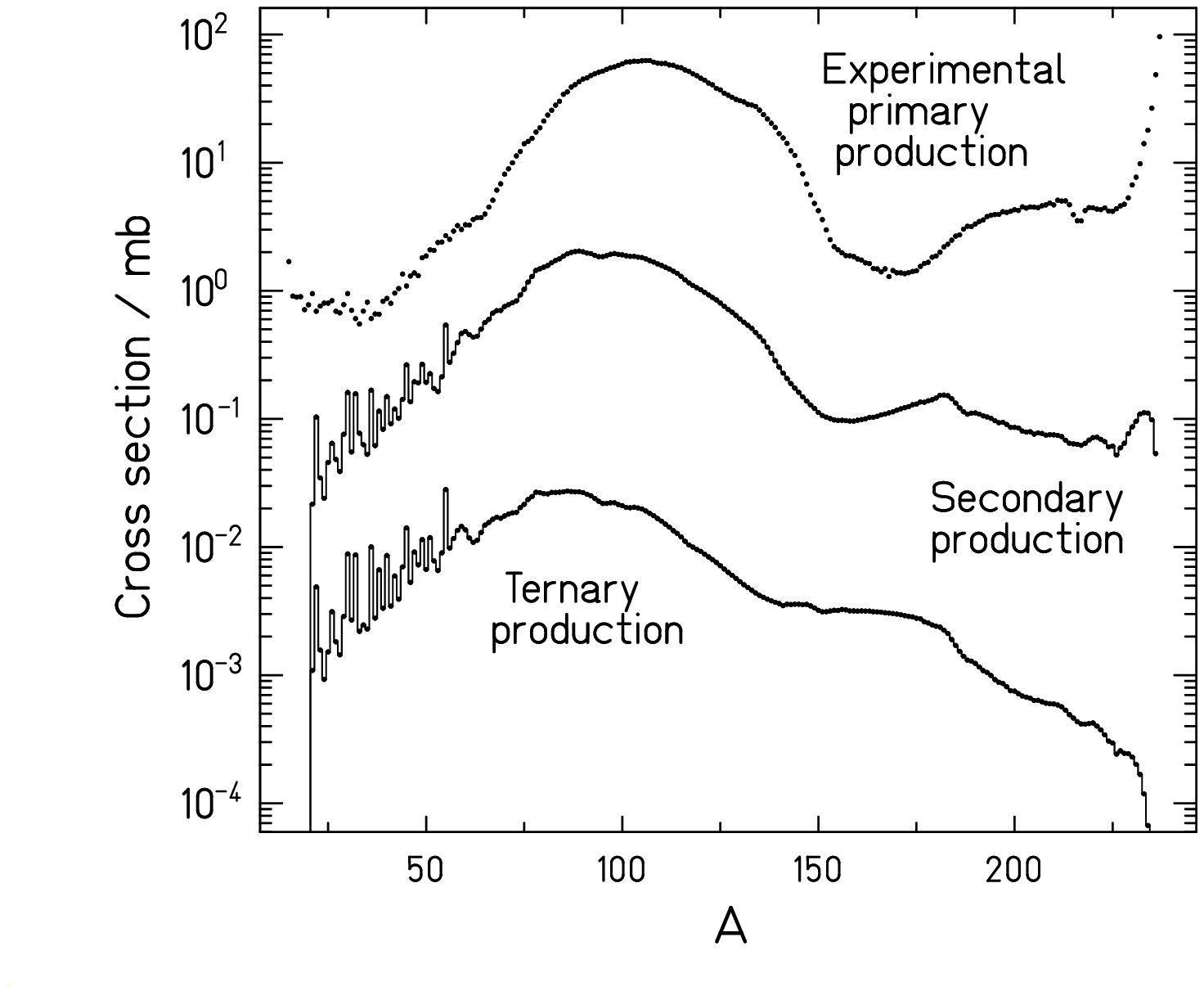,width=13cm}}
\caption{\label{fig:A}
Study of the evolution of the mass distribution of the residues 
generated in the reaction $^{238}$U+p at 1 A GeV for
different orders of multiple reactions in a hydrogen target
of thickness $87.3mg/cm^2$.  
The primary production corresponds to the 
experimental (dots) cross sections \cite{Bernas02,Taieb98,Ricciardi02}
corrected for the secondary contribution.
Nine elements from Tb to Ta are still not analysed 
and their cross sections are replaced by a calculation performed 
with the codes BURST \cite{Enqvist01} and 
ABLA-PROFI \cite{Gaimard91,Junghans98,Benlliure98} coupled together.
The secondary and ternary production is calculated according to 
Eq.~(\ref{eq:apparent}). The calculation has been performed using 
both the exact form (dots) of the term $\mathcal{A}_n$ 
and the approximation of equally-partitioned target (histogram)
described in Eq.~(\ref{eq:norderapprox}); the perfect overlapping
of the two calculations assures the reliability of the approximation.}
\end{center}
\end{figure}
The fundamental parameters to enter in the system of Eq.~(\ref{eq:apparent}) in
order to obtain a physical description of multiple reactions are the
production cross sections $\sigma_{i-1\to i}^{\sus \mathrm{evr}}$
and $\sigma_{i-1\to i}^{\sus \mathrm{ fis}}$. Due to the enormous number of these
parameters, a very fast numerical calculation is needed. The best solution we found 
is to couple a parameterisation to obtain the hot prefragments with a physical 
fission-evaporation code. 
\begin{figure}[b]
\begin{center}
\mbox{\epsfig{file=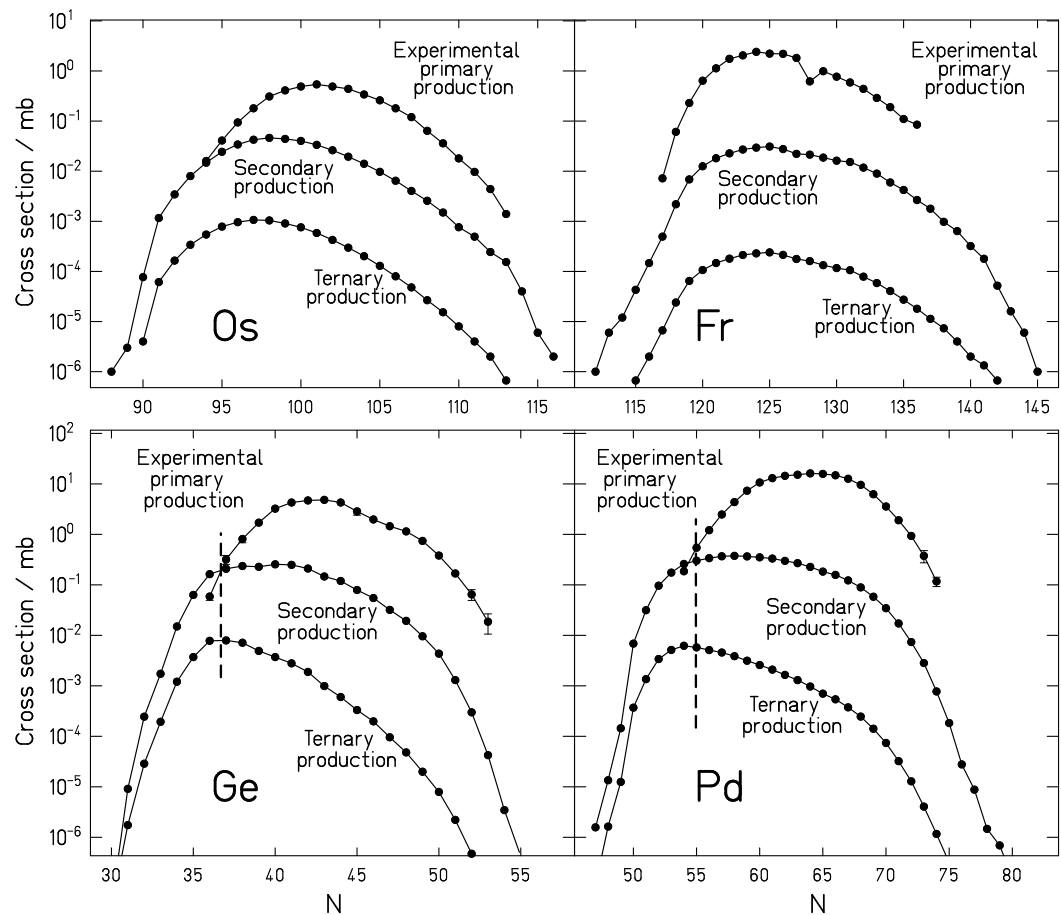,width=14cm}}
\caption{\label{fig:multiple}
Evolution of the isotopic distributions for different orders of multiple
reactions in the reaction $^{238}$U+p at 1 A GeV.
The different productions have been obtained as discussed 
in fig.~(\ref{fig:secondary}) and in fig.~(\ref{fig:A}). 
Evaporation residues (top): The mean neutron number of elements 
close to the projectile like Fr are not affected by consecutive reactions, 
while a shift towards lower neutron-numbers appears for elements like Os, 
produced in the light side of the mass-distribution slope.
Fission residues (bottom): The isotopic distribution of multiple-reaction
products is deformed and the mean neutron number tends to move towards the
residue corridor (dashed line).
}
\end{center}
\end{figure}
The routine BURST (applied for the correction for secondary reaction in the 
analysis presented in \cite{Enqvist01}) was chosen to reproduce the intra-nuclear cascade. 
The evaporation-stage has been simulated with the statistical
de-excitation code ABLA \cite{Gaimard91,Junghans98}
and the fission code PROFI \cite{Benlliure98}.
Also the possibility of multifragment emission, 
in the case of light and highly excited systems,
has been taken into account and simulated according to the reference
\cite{Schmidt02}.
The resulting isotopic production probability has then been normalized 
to the total reaction cross section obtained with the model of Karol-Brohm.

We verified that the results fit well with the available data of proton-reactions 
for $^{238}$U \cite{Bernas02,Taieb98,Ricciardi02} 
and $^{208}$Pb \cite{Enqvist01} at $1$ A GeV and $^{197}$Au 
\cite{Rejmund01,Benlliure01} at $800$ A MeV. 
In fig.~(\ref{fig:secondary}) the result of the calculation for $^{238}$U+p 
at $1$ A GeV is shown.
The primary production is represented by the experimental cross sections; 
due to the competition between fission and evaporation residues, nine elements 
(from Tb to Ta) have not yet been analysed; for the calculation, this 
gap has been filled with model-deduced cross sections.  
It should be noted that in this figure we anticipate the result on the experimental
primary production cross section, that have been disentangled from the secondary 
contribution by applying the procedure described in the following section. 
The secondary yields are one to two orders of magnitude lower than the primary 
production as shown in fig.~(\ref{fig:A}). 
The secondary evaporation residues are expected to extend 
to lighter masses and slightly overlap with the heavier fission products. 
In general, even if the 
mean neutron number of the isotopic distribution does not change significantly 
in respect to the primary evaporation residues, we observe an increased production of 
neutron-loss channels and an extension of the isotopic distribution towards 
the proton-rich side.
In contrast to the case of evaporation residues, the secondary fission residues are 
evidently less neutron rich. This is almost exclusively due to secondary 
evaporation residues that reduce the primary heavy neutron-rich fission 
products and populate the neutron deficient side; 
this is shown in fig.~(\ref{fig:multiple}).
The ternary production, three to four orders of magnitude lower than the 
primary production, shows an even more enhanced production around the residue
corridor.
The production resulting from secondary and ternary reactions, 
could be increased using a thicker target than the one considered in the 
present calculation.

\section{A recipe to disentangle primary and secondary reactions}
The calculation presented in fig.~(\ref{fig:secondary}) 
derives from the direct solution of the system of equations~(\ref{eq:apparent}),
applied to the measured reaction cross section of
$^{238}$U$+$p at $1$ A GeV.
Evidently, the result of the experiment was the cumulative
detection of the primary production together with the secondary,
without any clear experimental indication about how to 
select the primary yields.
Thus, the primary-reaction cross section were disentangled from the
secondary contribution solving the system~(\ref{eq:apparent}) 
inversely.

From the calculation shown in fig.~(\ref{fig:A}), we confirm
that the ternary contribution induced by a thin target  is even
lower than the uncertainties of the data
introduced by the measurement,
and it can be safely neglected.
In this case, the notation has been changed in respect to Eq.~(\ref{eq:apparent}).
The index ``fis,evr'' defines 
primary fission residues leading to secondary evaporation residues,
``evr,fis'' defines 
primary evaporation residues undergoing secondary fission 
and ``evr,evr''
primary evaporation residues producing secondary evaporation residues.
This consideration justifies the reduction of the
system~(\ref{eq:apparent}) to the second order of consecutive
reactions only.  
Thus, we can write the matrix relation
\begin{align}
	{\tilde {\cal X}}^{\sus\mathrm{evr}} &=
	{\cal L}^{\sus\mathrm{evr}} 
	{\cal X}^{\sus\mathrm{evr}} +
	{\cal G}^{\sus\mathrm{evr},\sus\mathrm{evr}}
	{\cal X}^{\sus\mathrm{evr}} 
	\label{eq:appfrag}
	\\
	{\tilde {\cal X}}^{\sus\mathrm{fis}} \;&=
	{\cal L}^{\sus\mathrm{fis}} 
	{\cal X}^{\sus\mathrm{fis}} +
	{\cal G}^{\sus\mathrm{fis},\sus\mathrm{evr}}
	{\cal X}^{\sus\mathrm{evr}} +
	{\cal G}^{\sus\mathrm{evr},\sus\mathrm{fis}} 
	{\cal X}^{\sus\mathrm{fis}} 
	\label{eq:appfiss}
	,
\end{align}
where $\tilde {\cal X}$ is the vector of apparent cross sections
directly measured by the experiment
and $\cal X$ is the set of the unknown primary cross sections
that we intend to extract:
\begin{gather}
	{\tilde {\cal X}} =
	\left( \begin{array}{c}
	{\tilde \sigma}_{(\uN_0\uZ_0\to \uN_0\uZ_0)}(\chi)	\\
	{\tilde \sigma}_{(\uN_0\uZ_0\to \uN_{0-1}\uZ_0)}(\chi)	\\
	\vdots				\\
	{\tilde \sigma}_{(\uN_0\uZ_0\to \uN\uZ)}(\chi)		\\
	\vdots				\\
	\end{array}\right)
	\quad
	,
	\quad 
	{\cal X} =
	\left( \begin{array}{c}
	\sigma_{(\uN_0\uZ_0\to \uN_0\uZ_0)}\\
	\sigma_{(\uN_0\uZ_0\to \uN_{0-1}\uZ_0)}	\\
	\vdots				\\
	\sigma_{(\uN_0\uZ_0\to \uN\uZ)}		\\
	\vdots				\\
	\end{array}\right)
	\notag
	. 
\end{gather}
The elements
\begin{gather}
	\alpha_{\uN\uZ} = 
	\frac{1}{2}\;
	{\cal A}_1(\uN_0\uZ_0,\uN\uZ,\chi)
	\notag\\
	\beta_{\uN_i\uZ_i} = 
	\frac{1}{2}\;
	\sigma_{(\uN_i\uZ_i\to \uN\uZ)} \chi \,
	{\cal A}_2(\uN_0\uZ_0,\uN_i\uZ_i,\uN\uZ,\chi) \,
	{\cal T}_2(\uN_i\uZ_i,\uN\uZ)
\label{eq:terms}
\end{gather}
are collected into the diagonal matrix ${\cal L}$ and the triangular matrix
${\cal G}$ respectively	:
\begin{gather}
	{\displaystyle\cal L} =
	\begin{scriptsize}
	\left( \begin{array}{ccccc}
	\alpha_{\uN_0\uZ_0}&&&&		\\
	&\alpha_{\uN_{0-1}\uZ_0}&&&	\\
	&&\ddots&&			\\
	&&&\alpha_{\uN\uZ}&		\\
	&&&&\ddots			\\
	\end{array}\right)
	\end{scriptsize}
	\;\;
	,
	\;\;
	{\cal G} =
	\begin{scriptsize}
	\left( \begin{array}{ccccccc}
	\beta_{\uN_0\uZ_0}&&&&&&			\\
	\beta_{\uN_0\uZ_0}&\beta_{\uN_{0-1}\uZ_0}&&&&&	\\
	\\
	\vdots&&&\ddots&&&				\\
	\\
	\beta_{\uN_0\uZ_0}&\ldots&\beta_{\uN,\uZ+1}
	&\ldots&\beta_{\uN+1,\uZ}&\beta_{\uN\uZ}&	\\
	\vdots&&&&&&\ddots				\\
	\end{array}\right)
	\end{scriptsize}
	\notag
	.
\end{gather}
The terms $\beta_{\uN_i\uZ_i}$
are evaluated for any of the involved secondary reaction 
through a model calculation of each
intermediate cross section $\sigma_{(\uN_i\uZ_i\to \uN\uZ)}$.
The attenuation terms ${\cal A}_1$ and ${\cal A}_2$, contained in 
$\alpha_{\uN\uZ}$ and in $\beta_{\uN_i\uZ_i}$, respectively,
are obtained by the calculation of the total reaction cross sections 
$\sigma_i$ through a model.
The transmission coefficient ratios ${\cal T}_2(\uN_i\uZ_i,\uN\uZ)$ are calculated
for any combination of intermediate residue $\uN_i\uZ_i$ and final residue
NZ as explained at the end of the present section.

The Eq.~(\ref{eq:appfrag}) describes the measured evaporation-residue 
yields as the sum of two contributions:
the quantity ${\cal L}^{\sus\mathrm{evr}}$
represents the {\it loss} of primary cross-section  due to the attenuation 
in the target. 
The term ${\cal G}^{\sus\mathrm{evr},\sus\mathrm{evr}}$ is the {\it gain} factor, which takes 
into account the increasing of the yields due to the residues produced
in two consecutive evaporation-residue steps.
The Eq.~(\ref{eq:appfiss}) refers to the fission-like yields, i.e. the
production-rate of those residues 
which the measurement attributes to fission events
according to the kinematical identification.
The apparent fission cross-section suffers from the attenuation in
the target, expressed by the term ${\cal L}^{\sus\mathrm{fis}}$, and
from two sources of gain: secondary evaporation residues of fission residues,
represented by 
${\cal G}^{\sus\mathrm{fis},\sus\mathrm{evr}}$ 
and evaporation residues followed by fission events, as described by
${\cal G}^{\sus\mathrm{evr},\sus\mathrm{fis}}$.
In the case of evaporation residues, the loss term depletes the yields of the
elements close to the projectile.
Part of these losses are redistributed by the gain factor 
${\cal G}^{\sus\mathrm{evr}}$
and populate the tail of light evaporation residues.
The effect of this process results into reducing the slope of the
mass distribution of the evaporation residues.
Another part of the losses goes into the term
${\cal G}^{\sus\mathrm{evr},\sus\mathrm{fis}}$,
estabilishing a coupling between the Eq.~(\ref{eq:appfrag}) and the 
Eq.~(\ref{eq:appfiss}) and populates the fission yields with secondary
fission fragments.
Since the primary evaporation residues that could undergo a secondary fission
event should be rather close to the projectile, the reaction products
should preserve essentially the structure of the primary fission
residues and do not introduce any modification or displacement in the
original primary distribution.
This is no more true when the primary reaction is a fission event.
In this case, the losses are redistributed by the gain factor
${\cal G}^{\sus\mathrm{fis},\sus\mathrm{evr}}$
as secondary evaporation residues of primary fission residues.
The effect of this process is a strong deformation of the fission-fragment
distribution. The primary fission products which, in general, are
neutron-rich are turned by secondary reactions into evaporation
residues, generally neutron deficient, and have the tendency to
cumulate around the residue corridor.
As a consequence, the whole fission distribution appears
less neutron-rich in the measurement than it is expected  
in reality. Moreover, the steeper is the slope of 
the fission distribution in the neutron-deficient side, 
the higher is the probability
that some residues are produced by secondary 
reactions in that region.

\begin{figure}
\noindent
\includegraphics[angle=90, height=0.96\textheight]{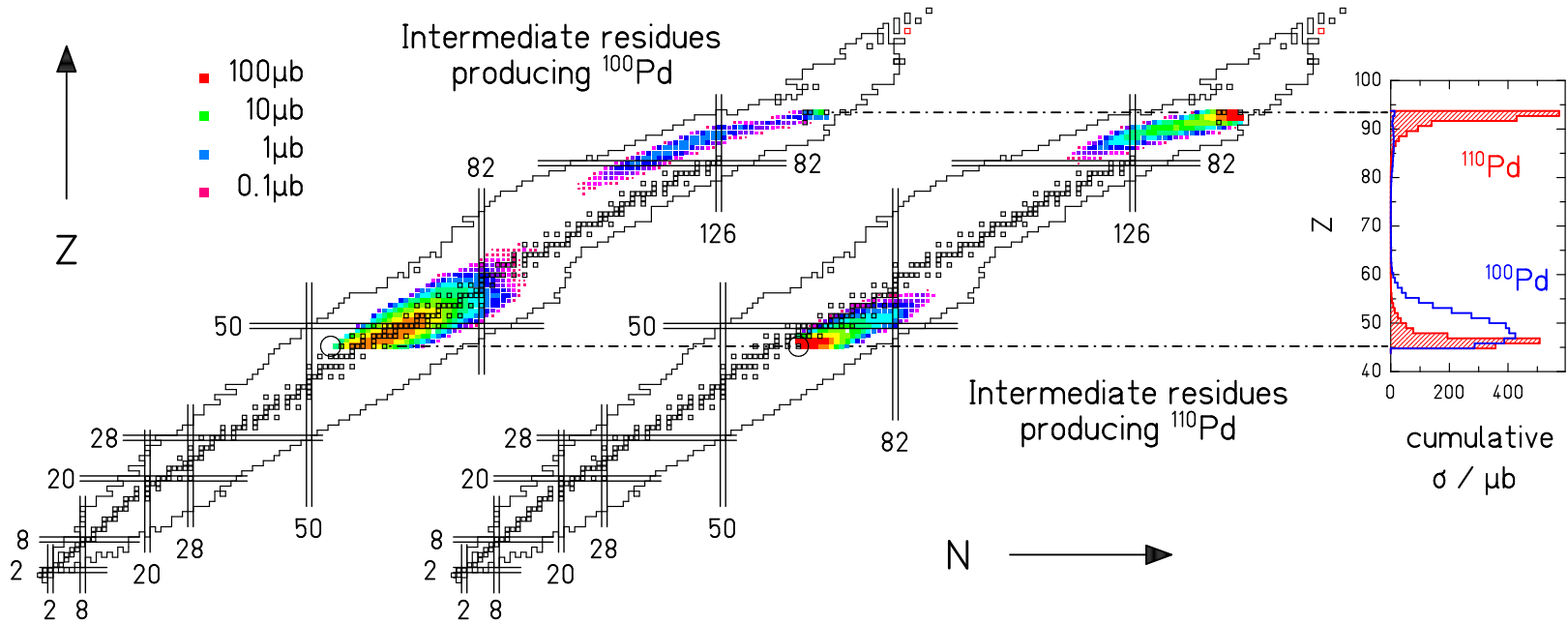}
\begin{sideways}
\parbox[h!]{0.96\textheight}
{
\caption[\textheight]{\label{fig:fathers}
Representation of the intermediate fragments $\uN_i\uZ_i$ that, 
once produced as residues of the reaction $^{238}$U+p at 1 A GeV, 
react a second time in the target and produce 
either $^{100}$Pd (left) or $^{110}$Pd (right).
On the isotopic chart we represent the secondary production cross sections 
for all the possible reactions
$\uN_i\uZ_i \to ^{100}$Pd or $\uN_i\uZ_i \to ^{110}$Pd,
multiplied by
$\sigma_{^{238}\mathrm{U} \to \uN_i\uZ_i}$
and divided by the total production cross section
for the primary residues of $^{238}$U.
As clearly shown in the Z-projection, while the secondary production
of neutron rich $^{110}$Pd can be equally attributed to fission 
or to evaporation residues of the neighbouring elements, the 
secondary production of the less neutron
rich $^{100}$Pd can be prevalently attributed to evaporation residues
of elements lighter than Nd.
}
}
\end{sideways}
\end{figure}

An example of the reaction mechanism ending up into the production of
a secondary fission residue is illustrated in fig.~(\ref{fig:fathers}),
where the isotopic distribution of the formation cross section 
of $^{110}$Pd and $^{100}$Pd for each possible intermediate mother 
nucleus is presented. 
We deduce that the stable and neutron-rich $^{110}$Pd 
is produced with high cross section mostly by primary fission of
heavy fragments, 
and its formation as a secondary residue 
should not enter in competition with its production by primary
reactions. 
Inversely, the secondary formation of $^{100}$Pd,
close to the residue corridor, results to depend almost exclusively
on secondary evaporation of the primary fission fragments:
the measured yield of formation of $^{100}$Pd should then be rather different from
its primary production cross-section.

Thanks to the triangularity of the matrix ${\cal G}$,
the system can be solved iteratively by the recurrence relation:
\begin{equation}
	\sigma_{(\uN_0\uZ_0\to \uN\uZ)}^{\mathrm{evr}} 
	\, \,= \,\,
	\frac{1}{\alpha_{\uN\uZ}^{evr}}
	\biggl(
	{\tilde{\sigma}}^{\mathrm{evr}}_{(\uN_0\uZ_0 \to \uN\uZ)}
	-\sum_{\uN_i\ge\uN \atop\uZ_i\ge\uZ }
	\sigma_{(\uN_0\uZ_0\to \uN_i\uZ_i)}^{\mathrm{evr}} 
	\beta_{\uN_i\uZ_i}^{evr}
	\biggr)
	.
\label{eq:matrixsolution}
\end{equation}
A similar equation holds for 
$\sigma_{(\uN_0\uZ_0\to \uN\uZ)}^{\mathrm{fis}}$.
Introducing the expressions ~(\ref{eq:terms}) in 
Eq.~(\ref{eq:matrixsolution}) and substituting the term ${\cal A}_2$
in the approximated form we obtain
the recursive relation:
\begin{equation}\begin{split}
	&
	\sigma_{(\uN_0\uZ_0\to \uN\uZ)}^{\sus\mathrm{evr}} 
	\quad=\quad
	{\tilde\sigma}^{\sus\mathrm{evr}}_{(\uN_0\uZ_0 \to \uN\uZ)}
	\,e^{\displaystyle\frac{\chi}{2}
	(\sigma_{\uN_0\uZ_0}+\sigma_{\uN\uZ})}
	+\\&\quad 
	-\frac{\chi}{2}
	\sum_{\uN_i\ge\uN \atop\uZ_i\ge\uZ }
	\biggl[
	\sigma_{(\uN_0\uZ_0\to \uN_i\uZ_i)}^{\sus\mathrm{evr}} 
	\sigma_{(\uN_i\uZ_i\to \uN\uZ)}^{\sus\mathrm{evr}}
	{\cal T}_2^{\sus\mathrm{evr,}\atop\sus\mathrm{evr}}(\uN_i\uZ_i,\uN\uZ)
	\,
	e^{\displaystyle\frac{\chi}{6}
	(\sigma_{\uN_0\uZ_0}-2\sigma_{\uN_i\uZ_i}+\sigma_{\uN\uZ})}
	\Biggr]
	\\
	&
	\sigma_{(\uN_0\uZ_0\to \uN\uZ)}^{\mathrm{fis}} 
	\quad=\quad
	{\tilde\sigma}^{\sus\mathrm{evr}}_{(\uN_0\uZ_0 \to \uN\uZ)}
	\,e^{\displaystyle\frac{\chi}{2}
	(\sigma_{\uN_0\uZ_0}+\sigma_{\uN\uZ})}
	+\\&\quad 
	-\frac{\chi}{2}
	\sum_{\uN_i\ge\uN \atop\uZ_i\ge\uZ }
	\Biggl\{\biggl[
	\sigma_{(\uN_0\uZ_0\to \uN_i\uZ_i)}^{\sus\mathrm{fis}} 
	\sigma_{(\uN_i\uZ_i\to \uN\uZ)}^{\sus\mathrm{evr}}
	{\cal T}_2^{\sus\mathrm{fis,}\atop\sus\mathrm{evr}}(\uN_i\uZ_i,\uN\uZ)
	\,+\\&\quad 
	+
	\sigma_{(\uN_0\uZ_0\to \uN_i\uZ_i)}^{\sus\mathrm{evr}} 
	\sigma_{(\uN_i\uZ_i\to \uN\uZ)}^{\sus\mathrm{fis}}
	{\cal T}_2^{\sus\mathrm{evr,}\atop\sus\mathrm{fis}}(\uN_i\uZ_i,\uN\uZ)
	\biggr]
	\,e^{\displaystyle\frac{\chi}{6}
	(\sigma_{\uN_0\uZ_0}-2\sigma_{\uN_i\uZ_i}+\sigma_{\uN\uZ})}
	\Biggr\}
	.
\label{eq:solution}
\end{split}\end{equation}
The relation~(\ref{eq:solution}) can be solved numerically for
decreasing masses of the observed fragments NZ. Following this order,
the unknown primary reaction cross section 
$\sigma_{(\uN_0\uZ_0\to \uN_i\uZ_i)}^{\sus\mathrm{evr,fis}}$
that appears in the system has been already calculated in
the previous step of the iteration.

\begin{figure*}[t]
\includegraphics[angle=-90,width=1\textwidth]{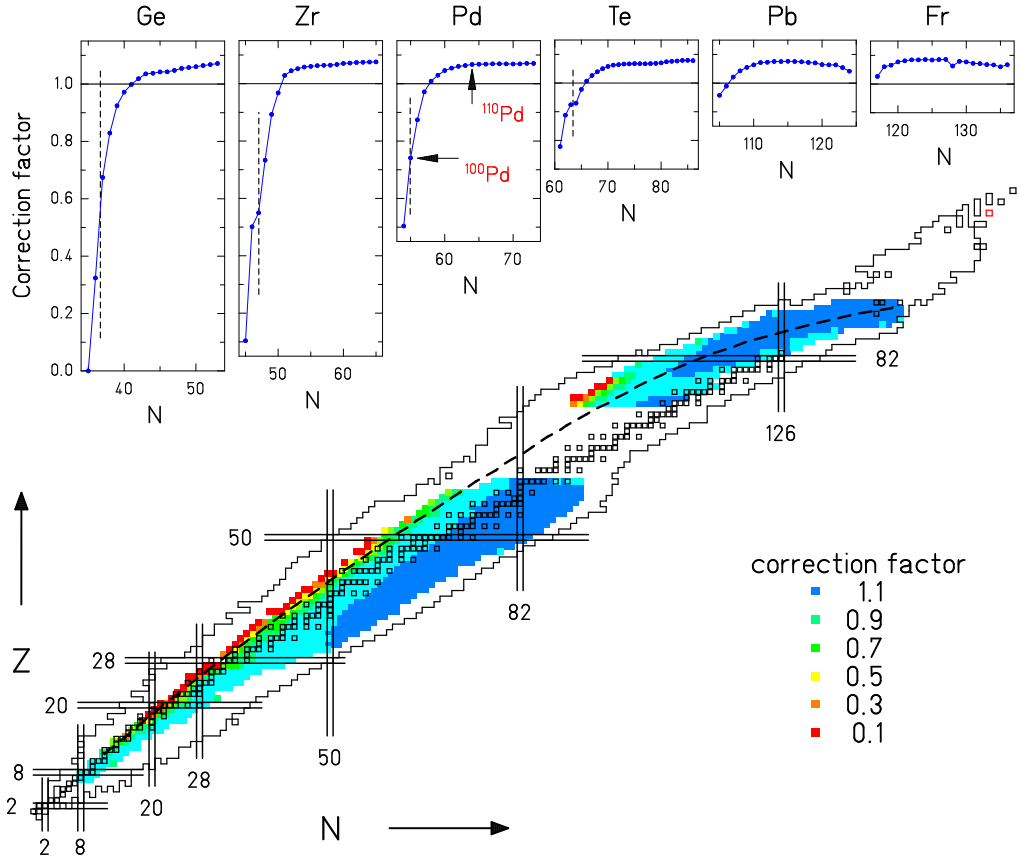}
\caption{\label{fig:hooks}
Bottom. Representation of the correction factor $h_{(\uN\uZ)}$ 
for secondary reaction on an isotopic chart for each residue 
measured in the reaction $^{238}$U+p at 1 A GeV
\cite{Bernas02,Taieb98,Ricciardi02}.
Top. Isotopic distributions of $h_{(\uN\uZ)}$ for the elements
Ge, Zr, Pd and Te, produced by fission and Pb and Fr, produced as 
evaporation residues.
The dashed lines represent the residue corridor, calculated
with the physical reaction code ABLA, imposing the sequential
evaporation of higly excited heavy prefragments.
The correction factor drops to very low values around the 
residue corridor: the steep slope of the ``hook'' shaped
$h_{(\uN\uZ)}$ distribution determines the limit for the 
observation of primary reaction products and for the
measurement of their isotopic cross sections.
The correction for $^{100}$Pd and $^{110}$Pd, indicated by 
arrows, can be compared to fig.~(\ref{fig:fathers}).
}
\end{figure*}
The solution of the system~(\ref{eq:solution}) gives directly the
value of the primary production cross-sections.
As discussed in the references \cite{Enqvist01,Benlliure01,Enqvist02}, 
a clear indication of the 
incidence of the secondary reactions on the
experimental measurement can be deduced studying the variation 
of the correction factor 
\[
	h_{(\uN\uZ)} = \sigma_{(\uN_0\uZ_0\to \uN\uZ)} \, / \,
		{\tilde\sigma}_{(\uN_0\uZ_0 \to \uN\uZ)}
\]
that we should apply to the apparent 
yields in order to obtain the primary cross-sections.
In fig.~(\ref{fig:hooks}) we mapped the values of $h_{(\uN\uZ)}$
for each measured isotope.
The distribution shows mainly a plateau for a positive 
correction of around 5\% to 7\%: these values,
originating from the first exponential term in Eq.~(\ref{eq:solution})
should compensate
the effect of the losses.
The homogeneity of the correction is then perturbed by the 
effect of the gain factor.
In some cases, like the evaporation-residue production, this effect does not
present complex features: generally, the secondary evaporation
residues with low mass-loss in respect to the projectile should 
distribute on the same ridge of the primary production
without showing a sensible deviation; as a consequence, the isobaric
distributions should simply be scaled in height according to a constant
factor and preserve the shape unchanged.
This prescription could be followed as a rather good approximation in
the case of heavy evaporation residues in reactions where no isotopes
with high fissility are produced: the correction $h_{(\uN\uZ)}$ would
reduce to a function of the mass-loss only. 
This was the approach applied in the analysis of the evaporation-residue 
production in the reaction of $^{197}$Au+p at $800$ MeV \cite{Rejmund01},
for the reaction $^{208}$Pb+p at $1$ GeV \cite{Enqvist01},
and for the reaction $^{208}$Pb+d at $1$ GeV \cite{Enqvist02}.
For these systems, and for the analysis of evaporation residues, 
this method would even have an advantage in respect to the 
application of the recursion~(\ref{eq:solution}): 
in fact,  the assumption of an
exclusive  dependence of the correction $h_{(\uN\uZ)}$ on the
mass-loss and the total cross-section of the intermediate
fragment $\sigma_{(\uN_i\uZ_i)}$ allows to build
a parameterisation based on the experimental data.
On the other hand, the relation~(\ref{eq:solution}) 
suffers from the uncertainties introduced by the theoretical 
calculation of the intermediate-fragment
cross-sections $\sigma_{(\uN_i\uZ_i\to \uN\uZ)}$.
Indeed, for systems like $^{238}$U, dominated by a
complicated competition between the fission and evaporation-residue channels,
the correction factor  $h_{(\uN\uZ)}$ can not be easily related to the
mass-loss. 
Moreover, the distribution of the evaporation residues of 
$^{238}$U+p widens considerably around Pb: as a consequence the rain
of secondary residues can not cover homogeneously the whole isobaric
distribution and it will populate prevalently the neutron deficient
side. As shown in fig.~(\ref{fig:hooks}) this effect appears 
in the neutron-distribution of the correction factor 
for the element Pb and it induces a slight reduction of 
$h_{(\uN\uZ)}$ for low neutron numbers.
This effect tends to disappear for heavier elements like Fr,
characterized by an almost constant correction factor.

The use of a detailed description of the secondary reaction mechanisms
becomes unavoidable when fission fragments should be treated: in 
this case $h_{(\uN\uZ)}$ largely changes as a function of the neutron
number and the relation~(\ref{eq:solution}) can not be substituted by
a parameterisation.
As firstly formulated and demonstrated in the analysis of the fission 
residues in the reactions $^{208}$Pb+p at $1$ GeV \cite{Enqvist01}, 
the correction factor is characterised by a ``hook'' shape and reduces steeply
when it approaches the residues corridor. However, 
the steepness of the ``hook'' reflects also the steepness of the
neutron-deficient side of the yield distribution.
In fig.~(\ref{fig:hooks}) an increased isotopic dependence of $h_{(\uN\uZ)}$ is
shown for decreasing element numbers.
The isotope $^{110}$Pd, whose formation mechanism has been illustrated
above, is on the plateau of the $h_{(\uN\uZ)}$ function and the
related cross-section should be corrected almost exclusively 
for the losses; on the contrary, $^{100}$Pd is located on the slope of
the correction factor.
Other elements like Ge or Zr show corrections that even approach 
100\% for their lightest isotopes: of course, 
when the correction is too large, the rejection of
these isotopes becomes necessary.
Isotopes showing correction factors smaller than $0.5$ are omitted
in the tabulation of cross sections in ref.~\cite{Bernas02}.
The steep slope of the correction factor becomes a technical limit
that excludes the possibility to extend the primary fission
cross-section measurement to the neutron-deficient side. 

\begin{figure}[b]
\begin{center}
\mbox{\epsfig{file=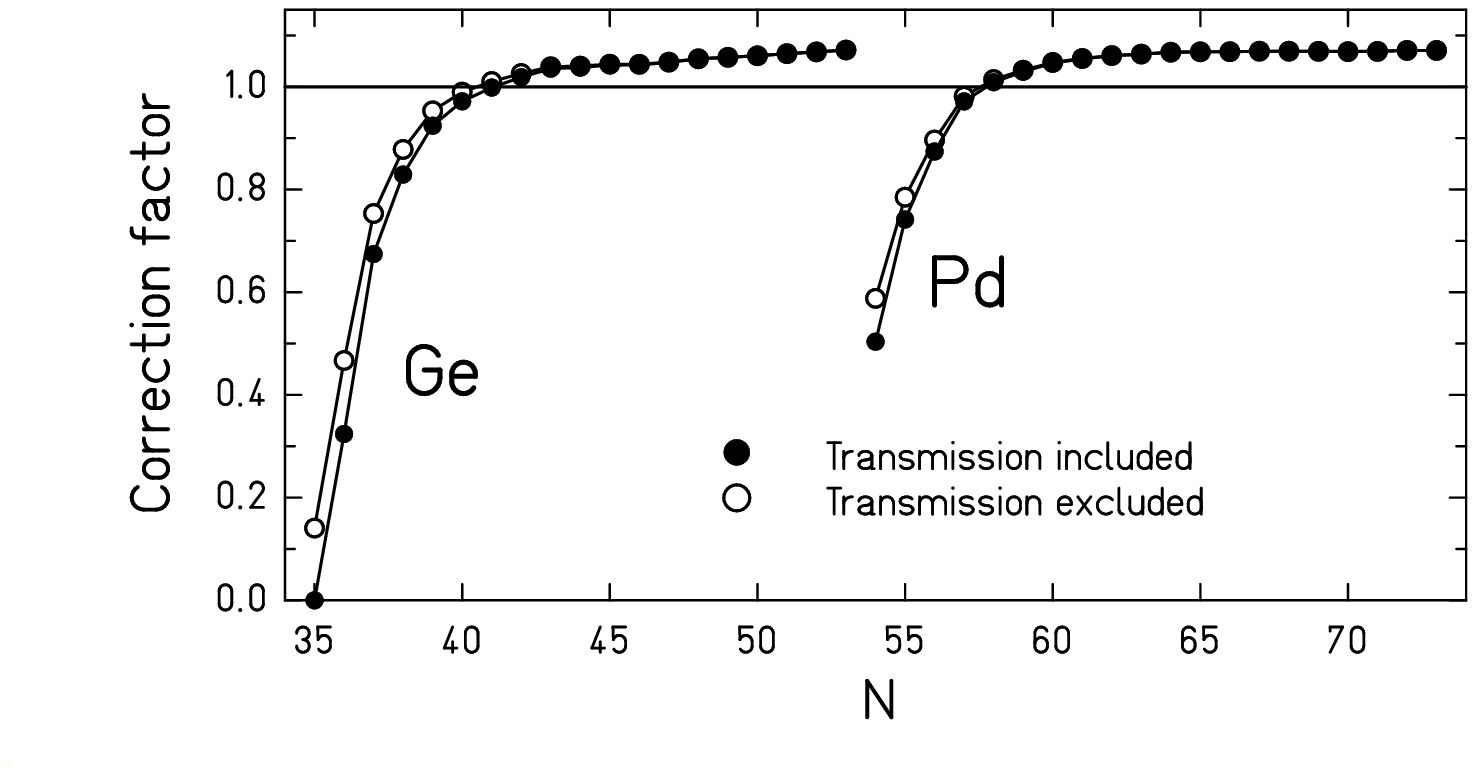,width=12cm}}
\caption{\label{fig:transmission}
Study of the contribution of transmission in the calculation 
of the correction factor $h_{(\uN\uZ)}$.
The transmission ratios are introduced as prescribed in
Eq.~(\ref{eq:tfrafra},\ref{eq:tfrafis},\ref{eq:tfisfra})}
\end{center}
\end{figure}

We should still point out that, as shown in the 
definition~(\ref{eq:terms}),
the gain factors include still another term:
the transmission ratio for the secondary fragments
${\cal T}_2(\uN_i\uZ_i,\uN\uZ)$
should be taken into account.
As shown in fig.~(\ref{fig:transmission}), the omission of the
transmission ratio leads to an overestimation of the correction
factor.
The geometry of the spectrometer, as discussed
in \cite{Benlliure02}, has to be evaluated precisely in order
to determine transmission values, which are deduced
relating the experimentally observed velocities
of the fragments, the acceptance of the spectrometer,
and the beam velocity.
The details of the calculation of these coefficients, used in this
work, is described in \cite{Bernas02}.
In the case of 
intermediate fragments $\uN_i\uZ_i$, that are not detected,
the transmission factor can not be measured directly.  
However, in this case we can still estimate the transmission factor 
because the velocities are known. 
In order to reduce the transmission ratios to
only experimental values it is worth while applying the following
approximations:
\begin{align}
	{\cal T}_2^{\sus \mathrm{evr},\sus \mathrm{evr}}(\uN_i\uZ_i,\uN\uZ) 
	\,&=\,
	\frac{t_2^{\sus \mathrm{evr},\sus \mathrm{evr}}
	(\uN_i\uZ_i,\uN\uZ)}
	{t^(\uN\uZ)}
	\, \approx \,
	\, 1     
	\label{eq:tfrafra}
	,\\
	{\cal T}_2^{\sus \mathrm{evr},\sus \mathrm{fis}}\,\,(\uN_i\uZ_i,\uN\uZ) 
	\,&=\,
	\frac{t_2^{\sus \mathrm{evr},\sus \mathrm{fis}}
	\,\,(\uN_i\uZ_i,\uN\uZ)}
	{t^(\uN\uZ)}
	\, \approx \,
	\, 1
	\label{eq:tfrafis}     
	,\\
	{\cal T}_2^{\sus \mathrm{fis},\sus \mathrm{evr}}\,\,(\uN_i\uZ_i,\uN\uZ) 
	\,&=\,
	\frac{t_2^{\sus \mathrm{fis},\sus \mathrm{evr}}
	\,\,(\uN_i\uZ_i,\uN\uZ)}
	{t^(\uN\uZ)}
	\, \approx \,
	\frac{t^{\sus \mathrm{fis}}(\uN_i\uZ_i)}
	{t^{\sus \mathrm{fis}}(\uN\uZ)}
	\,> 1     
	\label{eq:tfisfra}
	,
\end{align}
The relation~(\ref{eq:tfrafra}) follows from simple considerations 
on the angular distribution of evaporation residues.
In case of double reactions, the variance of the angular
distribution in the laboratory is simply the sum of individual
variances, which in turn are proportional to each individual
mass loss, implying that the total variance is proportional to the
total mass loss either for a secondary or primary reaction.
This leads to
$ 
t_2^{\sus \mathrm{evr,evr}}(\uN_i\uZ_i,\uN\uZ)\approx
	t^(\uN\uZ)
\, .
$
when the primary production of NZ is mainly related to evaporation residues.
We should observe that in some cases, when the primary production of NZ
is mainly related to fission, and there is an additional contribution 
coming from secondary evaporation residues of primary evaporation residues,
the ratio of Eq.~(\ref{eq:tfrafra}) becomes greater than the unit.
When fission is involved in the production of intermediate
or secondary fragments, even though the same addition rule 
for the variance is applied, the variance for a process leading to
evaporation residues 
is assumed to be much lower than in the case of fission, 
and it can be neglected. This leads to the relations
~(\ref{eq:tfrafis}) and
~(\ref{eq:tfisfra}). 
It should be observed that only the relation~(\ref{eq:tfisfra}) 
is relevant in the calculation
of the correction factor  $h_{(\uN\uZ)}$.
This derives from the observation that
the variance of the angular distribution of a fission fragment 
$\uN_i\uZ_i$ is not strongly modified when the residue reacts
again and produces a secondary residue by evaporation. 
We can then neglect the
contribution of the secondary reaction to the variance and substitute
the transmission coefficient 
$t_2^{\sus \mathrm{fis},\sus \mathrm{evr}}$,
that can not be calculated on the base of experimental observables,
with the transmission coefficient $t^{\sus \mathrm{fis}}$
of a fission fragment $\uN_i\uZ_i$; this leads to the the approximation
$
t_2^{\sus \mathrm{fis},\sus \mathrm{evr}}(\uN_i\uZ_i,\uN\uZ) 
\approx t^{\sus \mathrm{fis}}(\uN_i\uZ_i)
\, .
$	
Since the variance of the angular distribution for 
the fission fragment NZ is larger in respect to the lighter 
intermediate fission fragment $\uN_i\uZ_i$,
the resulting
term ${\cal T}_2^{\sus \mathrm{fis},\sus \mathrm{evr}}(\uN_i\uZ_i,\uN\uZ) $
is larger than the unit.

\section{Conclusion}

This work presents a study on the formalism 
of multiple nuclear reactions induced in a target.
A general model has been developed in order to
simulate the isotopic yields of secondary 
reactions, distinguishing between fission
and evaporation residues.
The model described in this work became the 
base of the numerical code ``SECONDARY''.
The program has been tested to reproduce 
the secondary-reaction correction used in the
analysis of the reactions 
$^{208}$Pb+p at $1$ A GeV \cite{Enqvist01}
.The comparison resulted in good agreement with 
the previous calculations.
The code SECONDARY, coupled with the reaction
code ABLA \cite{Gaimard91,Junghans98} and PROFI \cite{Benlliure98} 
has then been applied 
in the final step of the data-analysis 
presented in \cite{Bernas02},
aimed to determine
the isotopic cross section of the reaction 
$^{238}$U+p at $1$ A GeV.

\section{Acknowledgements }

We are endebted to Karl-Heinz Schmidt for 
his expertise in reaction-model calculations,
and for providing the codes BURST, ABLA and PROFI,
necessary for the calculations presented in this work. 
We are grateful to Orlin Yordanov for his interest 
and fruitful discussions.

\appendix
\section{Mathematical appendix }

\subsection{\protect\label{app:norder}Extension of the description 
of the secondary reactions to the $n^{th}$ order.}

The key relation to extend the description of the secondary reactions 
to the $n^{th}$ order is condensed in the following equality:
\begin{equation}
	\sum_{i=0}^{n-1}
	\frac{1}{
	\underset{\sss j=0\atop j\ne i}{\overset{\sss n}{\prod}}
	(\sigma_i-\sigma_j)}
	\quad=\quad
	-\frac{1}{
	\underset{\sss j=0}{\overset{\sss n-1}{\prod}}
	(\sigma_n-\sigma_j)}
	.
\label{eq:sumtoprod}
\end{equation}
Eq.~(\ref{eq:sumtoprod}) is a particular case, obtained imposing
$k=n-1$ and $\lambda=\sigma_n$, of the following more general form:
\begin{equation}
	\sum_{i=0}^{k}
	\frac{1}{
	(\sigma_i-\lambda)
	\underset{\sss j=0\atop j\ne i}{\overset{\sss k}{\prod}}
	(\sigma_i-\sigma_j)}
	\quad=\quad
	-\frac{1}{
	\underset{\sss j=0}{\overset{\sss k}{\prod}}
	(\lambda-\sigma_j)}
	.
\label{eq:sumtoprodgeneral}
\end{equation}
For $k=0$ the equality is trivial, and it can also be checked for  
$k=1$.
%
%
To prove Eq.~(\ref{eq:sumtoprodgeneral}) for any order, we should verify
that the expression is recursive.
If Eq.~(\ref{eq:sumtoprodgeneral}) is true at the order $k-1$, we can
extend it to the order $k$ with the following passages:
\begin{equation}\begin{split}
	&
	\displaystyle\sum_{i=0}^{k}
	\frac{1}{
	(\sigma_i-\lambda)
	\underset{\sss j=0\atop j\ne i}{\overset{\sss k}{\prod}}
	(\sigma_i-\sigma_j)} 
	\,=
	\\&=
	\displaystyle
	\frac{1}{\sigma_k-\lambda}
	\Biggl[
	\frac{1}{
	\underset{\sss j=0\atop j\ne i}{\overset{\sss k-1}{\prod}}
	(\sigma_k-\sigma_j)}
	+
	\sum_{i=0}^{k-1}
	\frac{\sigma_k-\lambda}{
	(\sigma_i-\lambda)
	\underset{\sss j=0\atop j\ne i}{\overset{\sss k}{\prod}}
	(\sigma_i-\sigma_j)}
	\Biggr] 
	\,=
	\\&=\,
	\displaystyle
	\frac{1}{\sigma_k-\lambda}
	\Biggl[
	\frac{1}{
	\underset{\sss j=0\atop j\ne i}{\overset{\sss k-1}{\prod}}
	(\sigma_k-\sigma_j)}
	+
	\sum_{i=0}^{k-1}
	\frac{\sigma_k-\lambda}
	{(\sigma_i-\sigma_k)(\sigma_i-\lambda)}
	\frac{1}{
	\underset{\sss j=0\atop j\ne i}{\overset{\sss k-1}{\prod}}
	(\sigma_i-\sigma_j)}
	\Biggr] 
	\,=
	\\&=\,
	\displaystyle
	\frac{1}{\sigma_k-\lambda}
	\Biggl[
	\frac{1}{
	\underset{\sss j=0\atop j\ne i}{\overset{\sss k-1}{\prod}}
	(\sigma_k-\sigma_j)}
	+
	\sum_{i=0}^{k-1}
	\frac{1}{
	(\sigma_i-\sigma_k)
	\underset{\sss j=0\atop j\ne i}{\overset{\sss k-1}{\prod}}
	(\sigma_i-\sigma_j)}
	-
	\sum_{i=0}^{k-1}
	\frac{1}{
	(\sigma_i-\lambda)
	\underset{\sss j=0\atop j\ne i}{\overset{\sss k-1}{\prod}}
	(\sigma_i-\sigma_j)}
	\Biggr]
	\notag
\end{split}\end{equation}
	or, applying Eq.~\ref{eq:sumtoprodgeneral} to 
	the last two terms in the brackets:
\begin{equation}\begin{split}
	&
	=\,
	\frac{1}{\sigma_k-\lambda}
	\Biggl[
	\frac{1}{
	\underset{\sss j=0\atop j\ne i}{\overset{\sss k-1}{\prod}}
	(\sigma_k-\sigma_j)}
	-
	\frac{1}{
	\underset{\sss j=0\atop j\ne i}{\overset{\sss k-1}{\prod}}
	(\sigma_k-\sigma_j)}
	+
	\frac{1}{
	\underset{\sss j=0}{\overset{\sss k-1}{\prod}}
	(\lambda-\sigma_j)}
	\Biggr] 
	\,=
	\qquad\qquad\quad
	\\&=\,
	-\frac{1}{
	\underset{\sss j=0\atop j\ne i}{\overset{\sss k}{\prod}}
	(\lambda-\sigma_j)}	
	.
\end{split}\end{equation}

%
The probability for $n$ recursive reactions in the target, expressed in the 
Eq.~(\ref{eq:norder}) reduces to the 
Eq.~(\ref{eq:0order},\ref{eq:1order},\ref{eq:2order})
for $n$ equal to 0,1, and 2, respectivly. We can demonstrate that it is
generally true if we obtain it recursively from 
${\cal P}_{n-1}(\uN_0\uZ_0,\uN_1\uZ_1,$ $\cdots$ $,\uN_{n-1}\uZ_{n-1},\chi)$
by applying Eq.~(\ref{eq:recursive}).
We reduce Eq.~(\ref{eq:norder}) to the order $n-1$ and we introduce it in 
Eq.~(\ref{eq:recursive}):
\begin{equation}\begin{split}
	&
	{\cal P}_n(\uN_0\uZ_0,\uN_1\uZ_1,\cdots,\uN_n\uZ_n,\chi) 
	\,= \\ &\qquad\qquad =\, 
	(-1)^{\dm n-1} 
	\,
	\sigma_{n-1\to n}
	\underset{\sss i=1}{\overset{\sss n-1}{\textstyle \prod}}
	\sigma_{(i-1\to i)}
	\,
	\sum_{i=0}^{n-1}
	\Biggl[
	e^{\dm -\sigma_n\chi}
	\int_{0}^{\chi}\ud \zeta
	\frac{e^{\dm (\sigma_n-\sigma_i)\zeta}}{
	\underset{\sss j=0\atop j\ne i}{\overset{\sss n-1}{\prod}}
	(\sigma_i-\sigma_j)}
	\Biggr]
	\\
	&\qquad\qquad =\, (-1)^{n-1}
	\, 
	\underset{\sss i=1}{\overset{\sss n}{\textstyle \prod}}
	\sigma_{(i-1\to i)}
	\, 
	\sum_{i=0}^{n-1}
	\frac{\, e^{\dm -\sigma_n\chi} - e^{\dm -\sigma_i\chi}}
	{
	\underset{\sss j=0\atop j\ne i}{\overset{\sss n}{\prod}}
	\dm (\sigma_i-\sigma_j)}
	.
	\qquad\qquad\quad
\label{eq:lastpassage}
\end{split}\end{equation}
If we apply the equality~\ref{eq:sumtoprod}
we can prove that the Eq.~(\ref{eq:lastpassage}) reduces to the form
(~\ref{eq:norder}).

\subsection{\protect\label{app:approx}Extension of the approximated 
secondary reaction formalism to the $n^{th}$ order.}

To prove that the Eq.~(\ref{eq:norderapprox}) is a recursive
extension of the Eq.~(\ref{eq:2orderapprox}) we should demonstrate
that it could be derived from the order $n-1$, i.e. we have to 
solve the integral
\begin{equation}
	\underset{\sss i=1}{\overset{\sss n}{\textstyle \prod}}
	\sigma_{(i-1\to i)}
	\,
	e^{\dm -\sigma_n\chi}
	\int_{0}^{\chi}
	\frac{(\zeta)^{n-1}}{(n-1)!}
	\,
	e^{\dm - \left(
	\frac{
	{\textstyle \sum_{j=0}^{n-1}}
	\, \sigma_j 
	}
	{n}
	-\sigma_n \right) \zeta}
	\ud \zeta	
	\nonumber
	.
\end{equation}
If, before integrating, we expand the exponential to the 
$1^{\mathrm{st}}$ order in 
$
	\frac{1}{n}\sum_{j=0}^{n-1}\, \sigma_j \zeta
	-\sigma_n \zeta
$, we reduce to the
approximated form:
\begin{equation}\begin{split}
	&
	\underset{\sss i=1}{\overset{\sss n}{\textstyle \prod}}
	\sigma_{(i-1\to i)}
	\,
	e^{\dm -\sigma_n\chi}
	\biggl[
	\frac{\chi^n}{n!}
	+\biggl(
	\frac{
	{\textstyle \sum_{j=0}^{n-1}}\, \sigma_j }
	{n}
	-\sigma_n
	\biggr)
	\frac{\chi^{n+1}}{(n-1)!(n+1)}
	\biggr]
	\,=
	\\&\quad\quad=\,
	\underset{\sss i=1}{\overset{\sss n}{\textstyle \prod}}
	\sigma_{(i-1\to i)}
	\,
	\frac{\chi^n}{n!}	
	\,
	e^{\dm -\sigma_n\chi}
	\,
	e^{\dm 
	-\biggl(
	\frac{
	{\textstyle \sum_{j=0}^{n-1}}\, \sigma_j }
	{n}
	-\sigma_n
	\biggr)
	\frac{n}{n+1}\,\chi
	}
	\notag
	,
\end{split}\end{equation}
where the resulting expression is equal to
Eq.~(\ref{eq:norderapprox}).


\begin{thebibliography}{99}
%
%
\bibitem{Bernas02} 
	M. Bernas, P. Armbruster, J. Benlliure, A. Boudard, E. Casarejos, S. Czajkowski, 
	T. Enqvist, R. Legrain, S. Leray, B. Mustapha, P. Napolitani, J. Pereira-Conca,
	F. Rejmund, M.V. Ricciardi, K.-H. Schmidt, C. St\'ephan, J. Taieb, L. Tassan-Got,
	C. Volant, 
	Submitted to {\it Nucl. Phys.} A.
%
\bibitem{Taieb98}
	J. Taieb, K.-H. Schmidt, L. Tassan-Got, P. Armbruster, J. Benlliure, M. Bernas, A. Boudard, 
	E. Casarejos, S. Czajkowski, T. Enqvist, R. Legrain, S. Leray, B. Mustapha, M. Pravikoff, 
	F. Rejmund, C: St\'ephan, C. Volant, W. Wlazlo
	submitted to Nucl. Phys. A,
	and thesis of J. Taieb, IPN-Orsay, 2000.
%
\bibitem{Ricciardi02}
	M.V. Ricciardi, K.-H. Schmidt, J. Benlliure, T. Enqvist, F. Rejmund, P. Armbruster, F. Ameil, 
	M. Bernas, A. Boudard, S. Czajkowski, R. Legrain, S. Leray, B. Mustapha, M. Pravikoff, 
	C. St\'ephan, L. Tassan-Got, C. Volant
	XXXIX Int. Winter Meeting on Nucl. Phys., Bormio, Italy (2001),
	and thesis in progress of M.V. Ricciardi, GSI-Darmstadt.
%
\bibitem{FRS} 
	H. Geissel, P. Armbruster, K.H. Behr, A. Br\"unle, K. Burkard, M. Chen, H. Folger, B. Franczak, 
	H. Keller, O. Klepper, B. Langenbeck, F. Nickel, E. Pfeng, M. Pf\"utzner, E. Roeckl, K. Rykaczewski, 
	I. Schall, D. Schardt, C. Scheidenberger, K.-H. Schmidt, A. Schroter, T. Schwab, K. S\"ummerer, 
	M. Weber, G. M\"unzenberg, T. Brohm, H.-G. Clerc, M. Fauerbach, J.-J. Gaimard, A. Grewe, E. Hanelt, 
	B. Kn\"odler, M. Steiner, B. Voss, J. Weckenmann, C. Ziegler, A. Magel, H. Wollnik, J.P. Dufour, 
	Y. Fujita, D.J. Vieira, B. Sherrill,
	{\it Nucl. Instrum. Methods} B {\bf 70}, {286} {(1992)}.
%
\bibitem{Karol75} 
	P. J. Karol,
	{\it Phys. Rev.} C {\bf 11}, {1203} {(1975)}.
%
\bibitem{Brohm94} 
	T. Brohm, K.-H. Schmidt,
	{\it Nucl. Phys.} A {\bf 569}, {821} {(1994)}.
%
\bibitem{Benesh89} 
	C.J. Benesh, B.C. Cook, J.P. Vary,
	{\it Phys. Rev.} C {\bf 40}, {1198} {(1989)}.
%
\bibitem{Enqvist01} 
	T. Enqvist, W. Wlazlo, P. Armbruster, J. Benlliure, M. Bernas, A. Boudard, S. Czajkowski, 
	R. Legrain, S. Leray, B. Mustapha, M. Pravikoff, F. Rejmund, K.-H. Schmidt, 
	C. St\'ephan, J. Taieb, L. Tassan-Got, 
	C. Volant, 
	{\it Nucl. Phys.} A {\bf 686}, {481} {(2001)}.
%
\bibitem{Gaimard91} 
	J.-J. Gaimard, K.-H. Schmidt, 
	{\it Nucl. Phys.} A {\bf 531}, {709} {(1991)}.
%
\bibitem{Junghans98} 
	A.R. Junghans, M. de Jong, H.-G. Clerc, A.V. Ignatyuk, G.A. Kudyaev,
	K.-H. Schmidt, 
	{\it Nucl. Phys.} A {\bf 629}, {635} {(1998)}.
%
\bibitem{Benlliure98} 
	J. Benlliure, A. Grewe, M. de Jong, K.-H. Schmidt, S. Zhdanov
	{\it Nucl. Phys.} A {\bf 628}, {458} {(1998)}.
%
\bibitem{Schmidt02}   
	K.-H. Schmidt, M.V. Ricciardi, A.S. Botvina, T. Enqvist,
	{\it Nucl. Phys.} A {\bf 710}, {157} {(2002)}.
%
\bibitem{Rejmund01} 
	F. Rejmund, B. Mustapha, P. Armbruster, J. Benlliure, M. Bernas, A. Boudard, J. P. Dufour, 
	T. Enqvist, R. Legrain, S. Leray, K.-H. Schmidt, C. St\'ephan, J. Taieb, L. Tassan-Got, C. Volant,
	{\it Nucl. Phys.} A {\bf 683}, {540} {(2001)},
	and thesis of B. Mustapha, IPN-Orsay, 1999.
%
\bibitem{Benlliure01} 
	J. Benlliure, P. Armbruster, M. Bernas, A. Boudard, J. P. Dufour, T. Enqvist, R. Legrain, 
	S. Leray, B. Mustapha, F. Rejmund, K.-H. Schmidt, C. St\'ephan, L. Tassan-Got, C. Volant,
	{\it Nucl. Phys.} A {\bf 683}, {513} {(2001)}.
%
\bibitem{Enqvist02} 
	T. Enqvist, P. Armbruster, J. Benlliure, M. Bernas, A. Boudard, S. Czajkowski, R. Legrain, S. Leray, 
	B. Mustapha, M. Pravikoff, F. Rejmund, K.-H. Schmidt, C. St\'ephan, J. Taieb, L. Tassan-Got, 
	F. Viv\`es, C. Volant, W. Wlazlo
	{\it Nucl. Phys.} A {\bf 703}, {435} {(2002)} 
%
\bibitem{Benlliure02} 
	J. Benlliure, J. Pereira-Conca, K.-H. Schmidt,
	{\it Nucl. Instrum. Methods.} A {\bf 478}, {493} {(2002)}.
\end{thebibliography}
\end{document}